\documentclass[useAMS,usenatbib]{mn2e}
\pdfoutput=1
\usepackage{graphicx} 
\usepackage{dcolumn}  
\usepackage{bm}       
\usepackage{amssymb,amsmath}  
\title[Clustering of Photometric Luminous Red Galaxies II: Cosmological Implications...]{Clustering of Photometric Luminous Red Galaxies II: Cosmological Implications from the Baryon Acoustic Scale}
\author[A. Carnero et al.]{A. Carnero$^1$, E. S\'anchez$^1$\thanks{E-mail:eusebio.sanchez@ciemat.es}, M. Crocce$^2$, A. Cabr\'e$^3$, E. Gazta\~naga$^2$\\
 $^1$Centro de Investigaciones En\'ergeticas, Medioambientales y Tecnol\'ogicas (CIEMAT), Madrid, Spain\\
 $^2$Institut de Ci\`encies de l'Espai (IEEC-CSIC), Barcelona, Spain \\
 $^3$University of Pennsylvania, Philadelphia, USA\\
}
\begin{document}

\date{\today}
\pagerange{1--7} \pubyear{2011}
\maketitle

\begin{abstract}
A new determination of the sound horizon scale in angular coordinates is 
presented. It makes use of $\sim 0.6 \times 10^6$ Luminous Red Galaxies, selected 
from the Sloan Digital Sky Survey imaging data, with photometric 
redshifts. The analysis covers a redshift interval that goes from $z=0.5$ to
$z=0.6$.  We find evidence of the Baryon Acoustic Oscillations (BAO) signal at 
the $\sim 2.3 \sigma$ confidence level, with a value of 
$\theta_{BAO} (z=0.55) = (3.90 \pm 0.38)^{\circ}$, including systematic errors. To our understanding, this is the first direct measurement 
of the angular BAO scale in the galaxy distribution, and it is in agreement with previous BAO measurements. We also show how radial 
determinations of the BAO scale can break the degeneracy in the measurement 
of cosmological parameters when they are combined with BAO angular measurements. The result is also in good agreement with the WMAP7 best-fit cosmology. We obtain a value of
$w_0 = -1.03 \pm 0.16$ for the equation of state parameter of the dark energy, or
$\Omega_M = 0.26 \pm 0.04$ for the matter density, when the other parameters are 
fixed. We have also tested the sensitivity 
of current BAO measurements to a time varying dark energy equation of 
state, finding $w_a = 0.06 \pm 0.22$ if we fix all the other parameters to the WMAP7 best-fit cosmology.
\end{abstract}

\section{Introduction}
\label{sec:intro}

The Baryon Acoustic Oscillations (BAO) signal imprinted in the galaxy
distribution was first convincingly detected in 2005 \citep{2005ApJ...633..560E}, using 
the Sloan Digital Sky 
Survey (SDSS)\footnote{http://www.sdss.org} data \citep{2000AJ....120.1579Y}. Since 
then, several more detections at different redshifts intervals have been 
done \citep{2007MNRAS.381.1053P,2009MNRAS.400.1643S,2009MNRAS.399.1663G,2010MNRAS.404...60R,2010MNRAS.401.2148P,2010ApJ...710.1444K}. All those used spectroscopic data in order to calculate galaxy redshifts, which 
has been the traditional technique to estimate distances, and measure 
the cosmological parameters through an averaged three-dimensional evaluation of
the BAO scale. 

Furthermore, the purely radial BAO 
signal was first obtained in \cite{2009MNRAS.399.1663G}, where they measure the 
redshift intervals that correspond to the sound horizon scale at two different redshifts. There 
are ongoing or proposed projects like BOSS \citep{2007AAS...21113229S} or 
BigBOSS \citep{2009arXiv0904.0468S} that will use massive measurements of galaxy 
spectra to obtain new measurements of the BAO scale, more precise and deeper than 
current determinations.

The feasibility of measuring the BAO signal using photometric redshifts (photoz) was first 
demonstrated in \cite{2007MNRAS.378..852P} (see also \cite{2007MNRAS.374.1527B} and 
\cite{2010arXiv1011.2448T}). Large 
multiband imaging surveys are a powerful 
way to explore the clustering properties of galaxies. Their advantage is the possibility
of mapping wider areas in deeper volumes, compared with spectroscopic surveys. However, the 
precision of the photoz is worse than its spectroscopic counterpart. There are 
projects like DES \citep{2005astro.ph.10346T}, PanSTARRS \citep{2000PASP..112..768K} or 
LSST \citep{2003NuPhS.124...21T} that aim to obtain the most precise cosmological constraints 
using imaging alone, compensating the lack of precision in redshift, with a larger volume and higher galaxy density, and also with the possibility of
studying different galaxy populations. 

In this paper, a new determination of the BAO angular scale is presented. The measurement has been done upon a photometric sample of Luminous Red Galaxies (LRG), obtained from the SDSS seventh data release (DR7),
published by \cite{2009ApJS..182..543A}. We study the BAO signal at a redshift interval that goes from $z=0.5$ to $z=0.6$, using the photometric sample 
described in the companion paper \citep{rsd_paper}. This measurement
is the first direct detection of the purely angular BAO signal, and it is complementary
to the purely radial BAO and to the three-dimensional averaged BAO signal, contributing to break degeneracies in the determination of cosmological parameters.

We use a model independent method to measure the angular BAO 
scale with precision \citep{2011MNRAS.411..277S}. The method is based on 
a parameterization of the angular correlation function as a sum of a power law, a constant 
and a Gaussian. There are a total of 6 parameters and the BAO position and its significance 
are given by the location and S/N in amplitude of the Gaussian. It is robust against systematic errors, it uses only observable quantities and it is independent of any assumption about other 
parameters or the shape of the correlation function. Furthermore, we have evaluated the systematic errors associated with the measurement.

The paper is organized as follows. First, we briefly explain the LRG selection criteria, that is common with the companion paper \citep{rsd_paper}. Then, the measurement 
of the angular BAO scale in the 2-pt angular correlation function is discussed. Next, we combine this result with previous measurements of the averaged and radial BAO scales to obtain new cosmological constraints, and finally, we comment on the 
conclusions.

\section{Selection of the Galaxy Sample}
\label{sec:sel}

\begin{figure}
\centering
\includegraphics[width=0.50\textwidth]{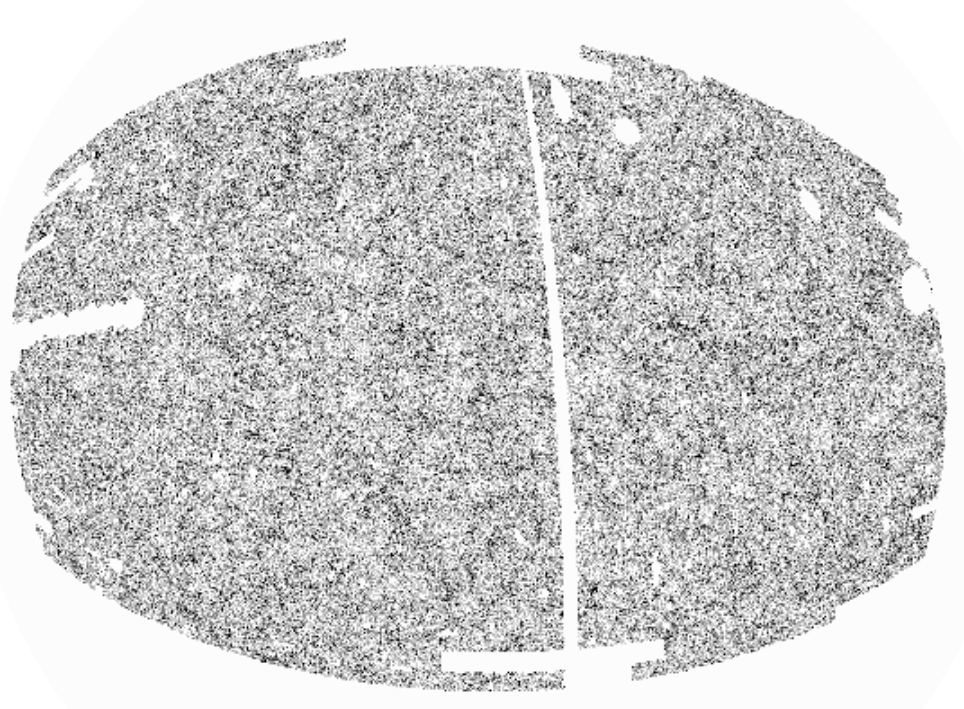}
\caption{The mask used in this analysis, as a Mollwide projection in equatorial coordinates. The white regions are excluded of the analysis. The vertical band is due to the photoz used in this analysis. \label{fig:mask}}
\end{figure}

We perform a color based selection of LRG in the SDSS-DR7 imaging sample, based on that published 
by \cite{2006MNRAS.372L..23C}. The selection is described in more detail in the companion 
paper \citep{rsd_paper}. First, we select the region in the color--color space
that is populated by LRGs \citep{2001AJ....122.2267E}, requiring 
$(r-i)>\frac{(g-r)}{4} + 0.36$ and $(g-r)>-0.72~(r-i) +1.7$. Then, we
minimize the star contamination, imposing an additional set of 
cuts: $17<r_{Petro}<21$, $0<\sigma_{r_{petro}}<0.5$, $0<(r-i)<2$, $0<(g-r)<3$ 
and $22 {\rm ~mag~ arcsec^{-2}} <\mu_{50}< 24.5 {\rm ~mag~arcsec^{-2}}$. In
these cuts, $g$, $r$ and $i$ are the model magnitudes corrected by extinction, $\sigma_{r_{Petro}}$ is 
the error on the Petrosian magnitude $r_{Petro}$ and $\mu_{50}$ is the mean surface brightness 
within the Petrosian half-light radius in the $r$ band. Finally, only galactic latitudes $b>25^{\circ}$
are considered. A total of $\sim 1.5 \times 10^{6}$ objects are selected, with 
a star contamination of $4 \pm 1\%$, determined using the spectroscopic SDSS sample.

We use the angular mask depicted in Figure \ref{fig:mask} as a Mollweide
projection in equatorial coordinates, which amounts a total area of 7136 square degrees. Only 
the largest contiguous area of the survey is considered in this analysis. There is also a 
vertical band in the mask that it is excluded due to the photoz catalog we have used, described 
in the next section. For further details, see \cite{rsd_paper}.

\section{Correlation Functions}
\label{sec:corrfunc}

\begin{figure}
\centering
\includegraphics[width=0.5\textwidth]{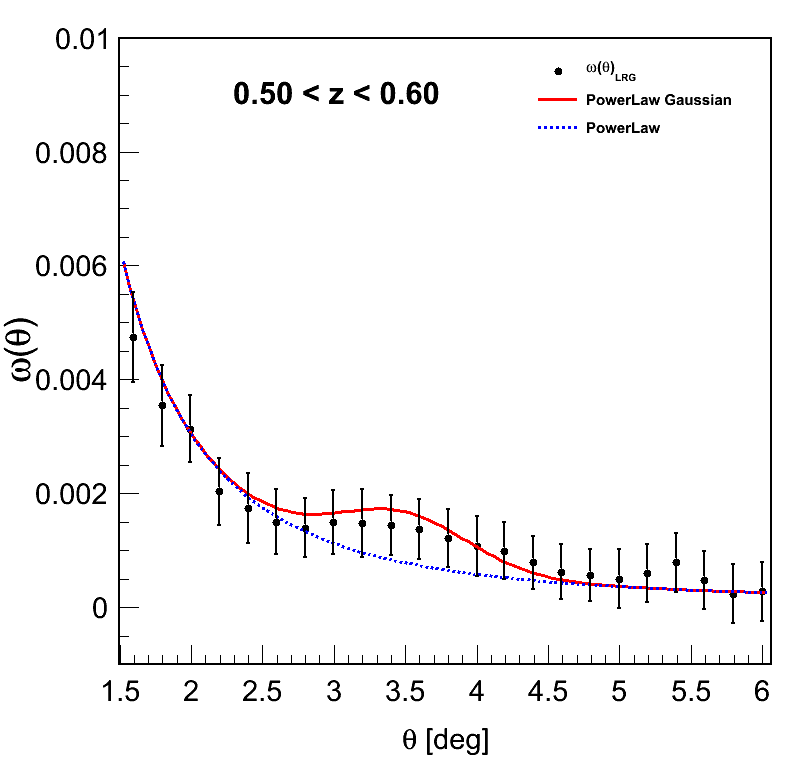} 
\caption{The observed angular correlation function for the analyzed redshift bin, 
         $\omega(\theta, z=0.55)$. The solid line is the result of fitting the 
         function to a power law plus a Gaussian, following
         the method presented in \protect \cite{2011MNRAS.411..277S}, and including the 
         covariance matrix. The dotted line shows only the power law. The BAO signal is
         detected with a statistical significance of $2.3 \sigma$. \label{fig:corr}}
\end{figure}

The photometric redshift determination developed in \cite{2008ApJ...674..768O} is
used in this analysis. It is combined with the value added catalog 
of \cite{2009MNRAS.396.2379C}, where the photometric redshift probability distribution 
function (PDF) for every object is given, following the method of \cite{2008MNRAS.390..118L}. We 
determine the redshift for each
object as the maximum of its PDF, but the full PDF distribution is used to estimate
the true distribution of objects with redshift. Objects
with PDF distributions that do not satisfy good quality requirements are 
rejected (see \cite{rsd_paper} for a detailed description of the photometric redshift
quality requirements). The highest redshift region, $0.5 < z < 0.6$, is chosen for this
analysis. It contains $0.61 \times 10^6$ objects.

\begin{figure}
\centering
\includegraphics[width=0.5\textwidth]{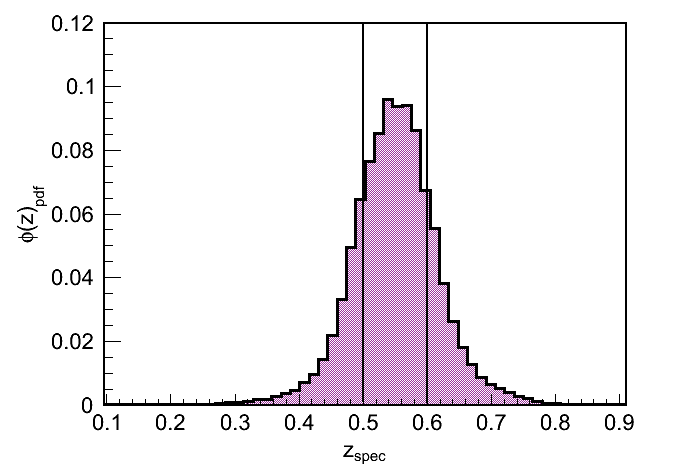} 
\caption{The true redshift distribution of selected galaxies. It has been obtained summing all the galaxies PDFs within the bin. The vertical lines are the limits of the photometric redshift bin under study.\label{fig:zbins}}
\end{figure}

We chose this bin width for two reasons. First, a much narrower bin does 
not enhance the sensitivity, due to the dilution effect caused by the 
photoz. And second, a much wider bin will decrease the amplitude of the correlation function, making more difficult to detect the BAO signal. 

Figure~\ref{fig:corr} shows  
the measured angular correlation function (dots) in this redshift bin. It has 
been computed using the Landy \& Szalay estimator \citep{1993ApJ...412...64L}.

The error and covariances in the angular correlation function have been 
evaluated using three independent methods. First, we have generated 50 artificial 
realizations of the observed $\omega(\theta)$. The error is computed as the dispersion of the generated 
realizations, taking into account the fraction of the sky that the survey 
covers. Second, we have used 
the standard Jack-Knife procedure, dividing the area in 80 regions. And third, we have 
used the theoretical description, following the result 
in \cite{2010arXiv1004.4640C}. The three determinations agree within
25\% in the square-root of the diagonal elements of the covariance matrix. To see a more detailed discussion of the error calculation, see \cite{rsd_paper}.

\section{Results}
\label{sec:results}

We use the method described in \cite{2011MNRAS.411..277S} to measure the 
BAO scale, using the observed correlation function of the previous 
section. The method is based on the description of $\omega(\theta)$ as the sum 
of a power law to describe the continuum and a Gaussian peak to describe the BAO
feature:

\begin{equation}
\label{eq:param}
\omega(\theta) = A + B \theta^{\gamma} + C e^{-(\theta - \theta_{FIT})^{2}/2\sigma^{2}}
\end{equation}

This expression is fitted to the data with free 
parameters $A$, $B$, $C$, $\gamma$, $\theta_{FIT}$ and $\sigma$. The true BAO scale
is recovered from the fitted parameter $\theta_{FIT}$ after correcting for the
projection effect due to the width of the redshift bin where the analysis is
performed:

\begin{equation}
\label{eq:corrtheta}
\theta_{BAO}=\alpha(z,\Delta z) \,\, \theta_{FIT}.
\end{equation}

The factor $\alpha$ is taken from the analysis presented 
in \cite{2011MNRAS.411..277S}, and depends on the redshift, $z$, and the 
redshift bin width, $\Delta z$.

The fit is presented in Figure \ref{fig:corr} (solid line), compared to a single power law (dotted line). This fit includes the full covariance
matrix, obtained with the Jack-Knife method, as seen in the companion paper. In \cite{rsd_paper} we calculate both the Jack-Knife and the analitical covariances, finding that both are compatible with each other and that results don't depend on the chosen covariance matrix \citep{rsd_paper}. 

The BAO peak is detected with a statistical significance of $2.3$ standard deviations. The statistical significance is computed using two methods. First, we use the fitted parameter $C$ from Equation~\ref{eq:param}, and we quote the significance as its difference from zero. Second, we compare the result with a theoretical model that includes the BAO wiggles and with a model that doesn't include them (this comparison is fully described in \cite{rsd_paper}). Both determinations agree in the significance of the detection.

\subsection{Systematic Errors}
\label{subsec:sys}

The main systematic errors affecting this measurement have been described 
in \cite{2011MNRAS.411..277S}, and we add them to our error budget here, except
for the photometric redshift, which is recalibrated. The reason for this 
recalibration is that the photometric redshift in SDSS is not perfectly 
Gaussian, as it was supposed in
\cite{2011MNRAS.411..277S}. 

We have recalibrated the photoz error by calculating the dispersion in the BAO scale if we compute $\omega(\theta)$ with different photoz codes, all contained in the SDSS DR7 catalog. We redo the full analysis using
the photometric redshift given by the \textit{photozd1} and also by using the full PDF as the redshift, and not only the maximum probability. In the latter case, all galaxies has a probability to lay within the selected redshift bin, for example, in the redshift bin $0.5\leq z \leq 0.6$. The only difference in the numerical algorithm is that, instead of calculating distances between pairs of galaxies, now we calculate distances between probabilities of galaxies, and weight the distance depending on the probability of pairs in the given bin.

We found that the amplitude of the correlation function does change, being smoother if we use the full PDF, but the measured scale $\theta_{BAO}$ is stable within 2\% for all cases, including the \textit{photozd1} estimation. Therefore, we add a contribution of 2\%, coming from the different photoz code used, to the 5\% found in \cite{2011MNRAS.411..277S}, which only accounts for the smearing of the photoz in a ideal Gaussian case. The total systematic error due to photometric redshift is then $\Delta \theta_{BAO} (photoz) = 7\%$.


Moreover, there are some additional effects that need to be taken into account on top
of those of \cite{2011MNRAS.411..277S}. These are the effects 
related to the selection of the sample (selection cuts, mask and contamination).

To evaluate the systematic error associated to the selection criterias, we vary the selection cuts of section \ref{sec:sel} and recompute $\omega(\theta)$. Then, we compare the new measured $\theta_{BAO}$ with the nominal value. We have tested variations in $\mu_{50}$, $(r-i)$ and $(g-r)$ within 10\% of their nominal limits in section~\ref{sec:sel}. The maximum variation found is $\Delta \theta_{BAO}(selection)=2.5\%$, when we vary the $\mu_{50}$ limits. In all cases, variations in $\theta_{BAO}$ are always below the statistical uncertainty, that in our case is $\sim 5\%$. $\Delta \theta_{BAO}(selection)=2.5\%$ is assigned as the systematic error due to the sample selection. This is actually likely to be conservative and most likely larger than the true $1\sigma$ value, but it is almost negligible compared to the 7\% uncertainty of the photoz and the 5\% statistical uncertainty.

We have also varied the mask in order to assess further uncertainties in our result. First we relax the cut in $b > 25^{\circ}$, and allow areas with $b < 25^{\circ}$. The effect of this cut in the galaxy number density can be seen in Figure 3 of the companion paper \cite{rsd_paper}. There is a gradient in the number density of galaxies at low galactic latitudes, implying there is a population of objects, belonging to our galaxy, that is contaminating the sample. The effect in $\theta_{BAO}$ is similar to the effect that we obtain when we vary the galaxy selection cuts. Both produce smooth distortions in the clustering signal.

We also compute $\omega(\theta)$ for different quality values of extinction, seen at the bottom panel of Figure 4 in the companion paper \cite{rsd_paper}. We eliminate areas with extinction higher than 0.15, 0.2 and without limits. Variations in $\theta_{BAO}$ are almost negligible, as we are dominated by the statistical error. For the mask, the strongest difference occurs if we allow areas with $b > 20^{\circ}$. The variation is of the 2\%, associated with the mask systematic error as $\Delta \theta_{BAO} (mask) = 2\%$.

We recall that both error estimations are likely to be overestimated, althought both are almost negligible in comparison with the statistical error of 5\% and the 7\% photoz systematic error, therefore, and considering the level of precision in the data, we believe it is reasonable to follow this prescription.

The influence of stars contamination has also been checked. We have fitted the
correlation function including the star contamination as a constant factor, 
multiplied by the star correlation function (see \cite{rsd_paper}). The influence 
is found to be negligible, since the correlation function of stars 
varies very slowly with the angular scale. Variations in $\theta_{BAO}$ are $\Delta \theta_{BAO}(stars)<0.1\%$. Moreover, we have included
different levels of stars contamination in the theoretical
correlation functions presented in \cite{2011MNRAS.411..277S}, always as a constant factor 
multiplied by the star correlation function. We have applied our method to all of 
them and we found that the $\theta_{BAO}$ obtained with our parameterization 
is insensitive to the presence of stars, at least for contaminations up to 10\%.

\begin{figure}
\centering
\includegraphics[width=0.5\textwidth]{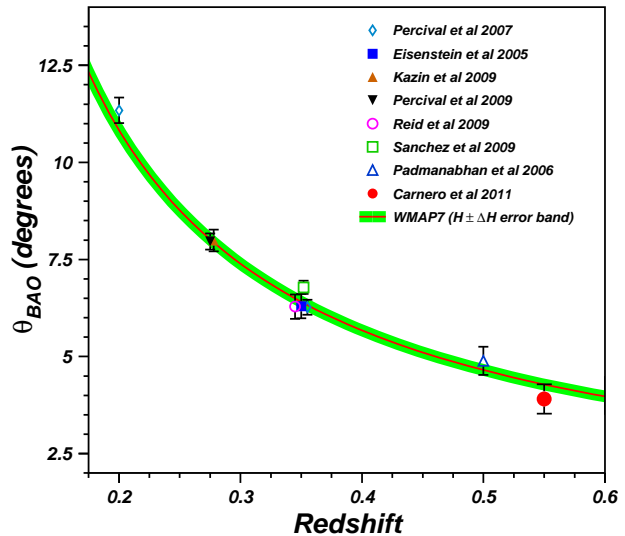} 
\vspace{-1.0truecm}
\caption{The angular BAO scale as a function of redshift. The measurement presented in this paper is the solid dot. Its error includes both the statistical and the systematic contributions. The other measurements are inferred from the given reference by translating the three-dimensional averaged distance parameter $D_V$, into an angular scale, with the fiducial cosmology used in each reference. All of them are compatible with the WMAP7 cosmology (solid line). The band includes the uncertainty in the Hubble constant $H_0$. Some of the measurements at $z=0.35$ have been slightly displaced in redshift for clarity.\label{fig:baovsz}}
\end{figure}

The full accounting of systematic errors is presented in Table~\ref{tab:sys}. The
total systematic error budget is $\Delta \theta_{BAO} (sys) = 8.0\%$.

About the significance of the detection, we found that the effect of these systematics in the significance can produces maximum changes of the order of the $10\%$, up to a significance of $2.6 \sigma$, if we cut in $b>20^{\circ}$ and using the maximum probability as the redshift and the nominal color cuts. We don't find any clear correlation with cuts or with which photoz code we used. This is due to the fact that we are yet limited by statistical uncertainties and the smearing of the photoz.

\begin{figure*}
\centering
\leavevmode
\begin{tabular}{cc}
\includegraphics[width=0.5\textwidth]{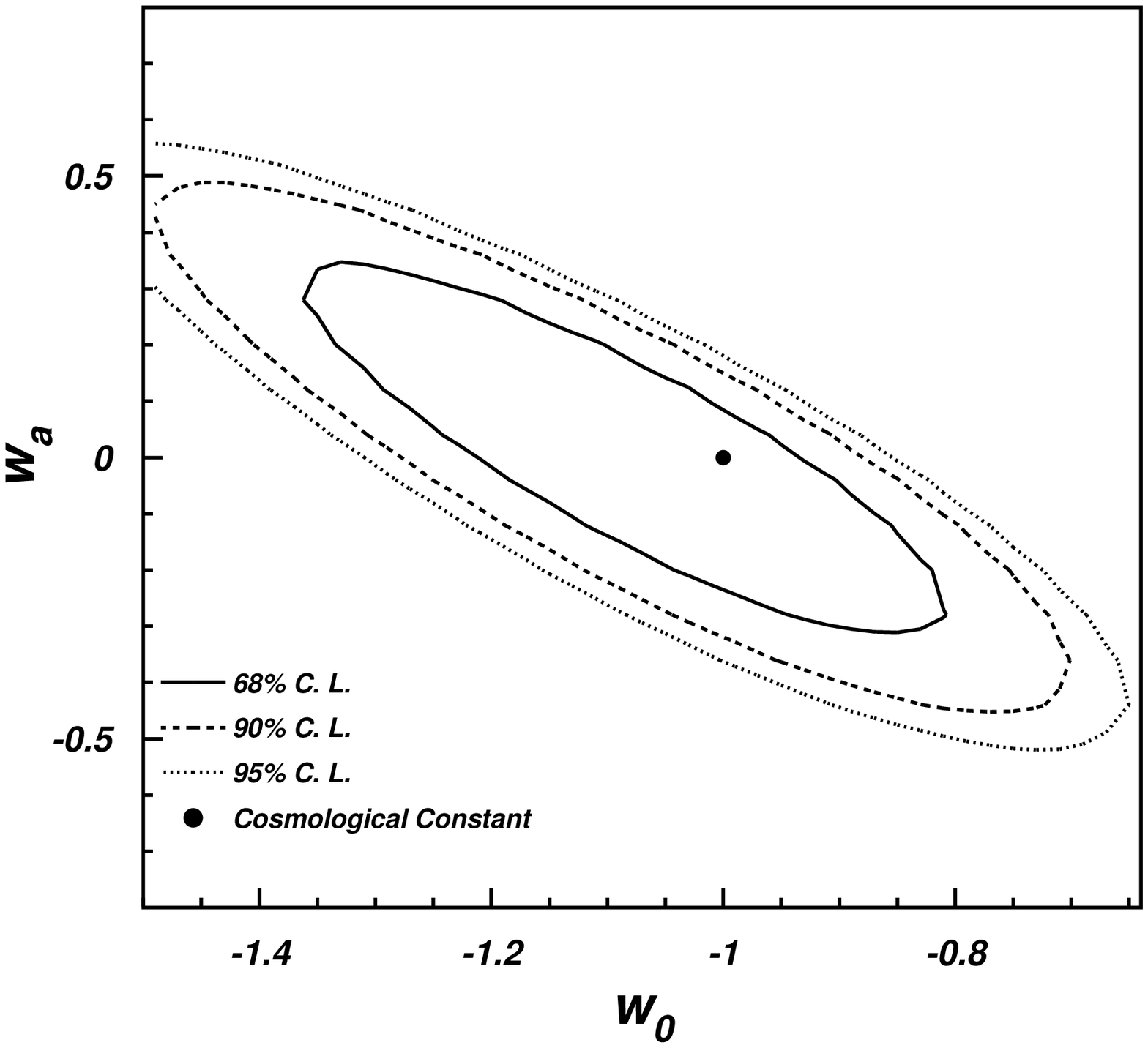} &
\includegraphics[width=0.5\textwidth]{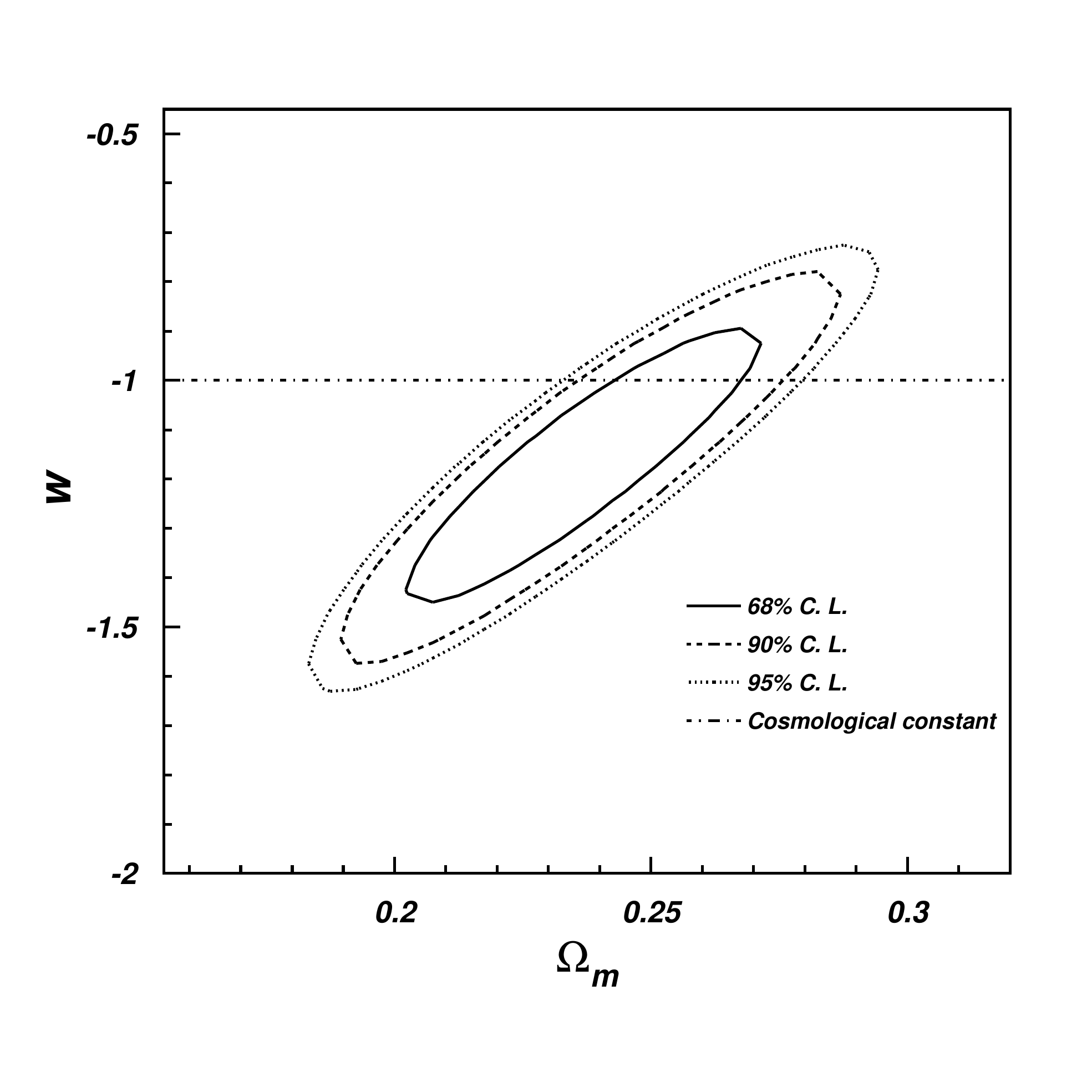}  \\
\end{tabular}
\vspace{-1.0truecm}
\caption{Constraints in the equation of state of the dark energy using the data in Figure~\ref{fig:baovsz} and the radial BAO measurements from \protect \cite{2009MNRAS.399.1663G}. Left: Constraints 
         in the plane $w_{0}$, $w_{a}$, where $w(a) = w_{0} + w_{a} (1-a)$, and $a$ the scale factor 
         of the universe. Data is compatible with a dark energy that behaves as a cosmological constant to better than $1\sigma$. 
         (solid dot). Right: Constraints in the plane $w$, $\Omega_{m}$. The dash-dotted line corresponds to the Cosmological Constant and is compatible with the data.\label{fig:w0wa}}
\end{figure*}

\begin{table}
\centering
\begin{tabular}{ccc}
\hline
Systematic error & $\Delta \theta_{BAO}$ \\
\hline\hline
Parameterization                & 1\% \\
Photometric redshift error     & 7\% \\
Redshift space distortions     & 1\% \\
Theory                         & 1\% \\
Projection effect              & 1\% \\
\hline
Selection of the Sample        & 2.5\% \\
Survey mask                    & 2\% \\
\hline
\end{tabular}
\caption{Estimation of systematic errors in the determination of $\theta_{BAO}$. Those above the 
horizontal line are taken from \protect \cite{2011MNRAS.411..277S}, except the error coming from the
photometric redshift, which has been recalibrated for this work. The others are the main new 
contributions that arise from a real photometric redshift survey.\label{tab:sys}}
\end{table}

\subsection{Comparison with Previous Measurements}
\label{subsec:fullresult}

The fitted angular scale, associated with the BAO scale is $\theta_{FIT}(z=0.55) = 3.48 \pm 0.19 ^{\circ}$, obtained upon a redshift bin of width $\Delta z_{photo}=0.1$. Following \cite{2011MNRAS.411..277S}, this
value must be corrected to obtain the true BAO angular 
scale, $\theta_{BAO} = \alpha(z,\Delta z) \theta_{FIT}$. For 
this redshift and bin width, the correction is 
$\alpha(z=0.55,\Delta z_{true}= \sqrt{2\pi} \Delta z_{photo}) = 1.12$. The 
final result for the BAO scale, including statistical and systematic
errors is $\theta_{BAO}(z=0.55) = 3.90 \pm 0.38 ^{\circ}$.

The evolution of the angular BAO scale with redshift is presented in 
Figure~\ref{fig:baovsz}. The measurement presented in this analysis is the solid dot, including the statistical and the systematic errors. The other measurements 
are inferred from the given reference, by translating the three-dimensional averaged 
distance parameter $D_V$ into an angular scale, with the fiducial cosmology used in 
each reference. All the measurements are in agreement with the WMAP7 cosmology to better than $1\sigma$, represented by the solid line. The band includes the uncertainty in the determination of the Hubble constant $H_0$ \citep{2011ApJ...730..119R}.

We have tested the sensitivity of current BAO measurements to a dark energy with an equation of state parameter that varies with time, as $w(a) = w_{0} + w_{a} (1-a)$, where $a$ is the scale factor of the universe. In Figure~\ref{fig:w0wa} (Left) we show the allowed regions for parameters  $w_{0}$ and $w_{a}$, when we fit the evolution of the BAO scale using the data in Figure~\ref{fig:baovsz}, plus the radial BAO scales at 
$z=0.24$ and $z=0.43$ from \cite{2009MNRAS.399.1663G}, which values are $r_{s}(z=0.24)=110.3\pm 2.9\pm 1.8 ~ Mpc/h$ and $r_{s}(z=0.43)=108.9\pm 3.9 \pm 2.1 ~ Mpc/h$. We fix all the other cosmological parameters to their values in the WMAP7 $w$CDM cosmology~\citep{2011ApJS..192...18K}, and we use the most recent Hubble constant determination \citep{2011ApJ...730..119R}. We have chosen
the values of \cite{2007MNRAS.381.1053P} at $z=0.20$, \cite{2010MNRAS.401.2148P} at 
$z=0.275$, \cite{2009MNRAS.400.1643S} at $z=0.35$ and
the measurement presented in this paper at $z=0.55$, since the other measurements at the
same or similar redshifts are strongly correlated. Data is compatible at the $1\sigma$ level with a dark energy that behaves as a cosmological constant  ($w_0 = -1$, $w_a = 0$, included as the solid dot in Figure~\ref{fig:w0wa}).

We have also studied the sensitivity of current BAO measurements in the 
plane $w$, $\Omega_{m}$, under the same conditions than above: we use the same data 
points and also include the radial BAOs. The
other parameters remain fixed plus $w_{a}$, which is fixed to zero. The
result is presented in Figure~\ref{fig:w0wa} (Right). The dash-dotted line
shows the expected value for the Cosmological Constant, compatible with the data. 

If we fit to a single parameter, it gives $w = -1.03 \pm 0.16$. The data is
compatible with a cosmological constant with a precision of the $\sim 20\%$ in $w$. If we do the same
study for $\Omega_M$, we find a value of $\Omega_M = 0.26 \pm 0.04$. Both measurements
include systematic errors, and have been obtained fixing all the other parameters to
the same values than above, except $H_0$, which has been varied within its uncertainty. We have now included the influence of $H_0$ because it is larger in this case. The error budget in the parameter becomes 40\% larger, including the $H_0$
uncertainty. The influence in the previous 2-dimensional constraints is smaller.

\begin{figure*}
\centering
\leavevmode
\begin{tabular}{cc}
\includegraphics[width=0.5\textwidth]{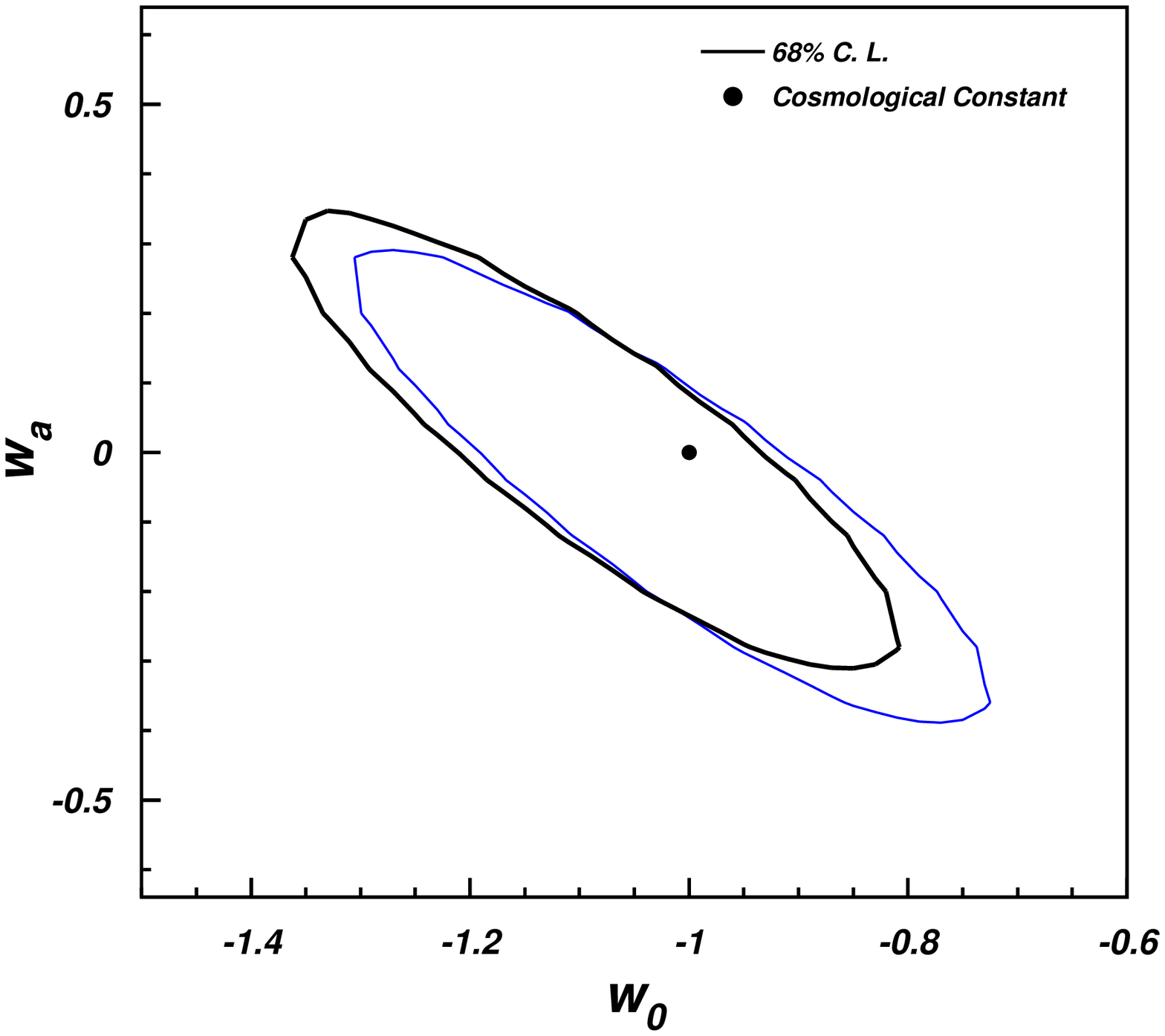} &
\includegraphics[width=0.5\textwidth]{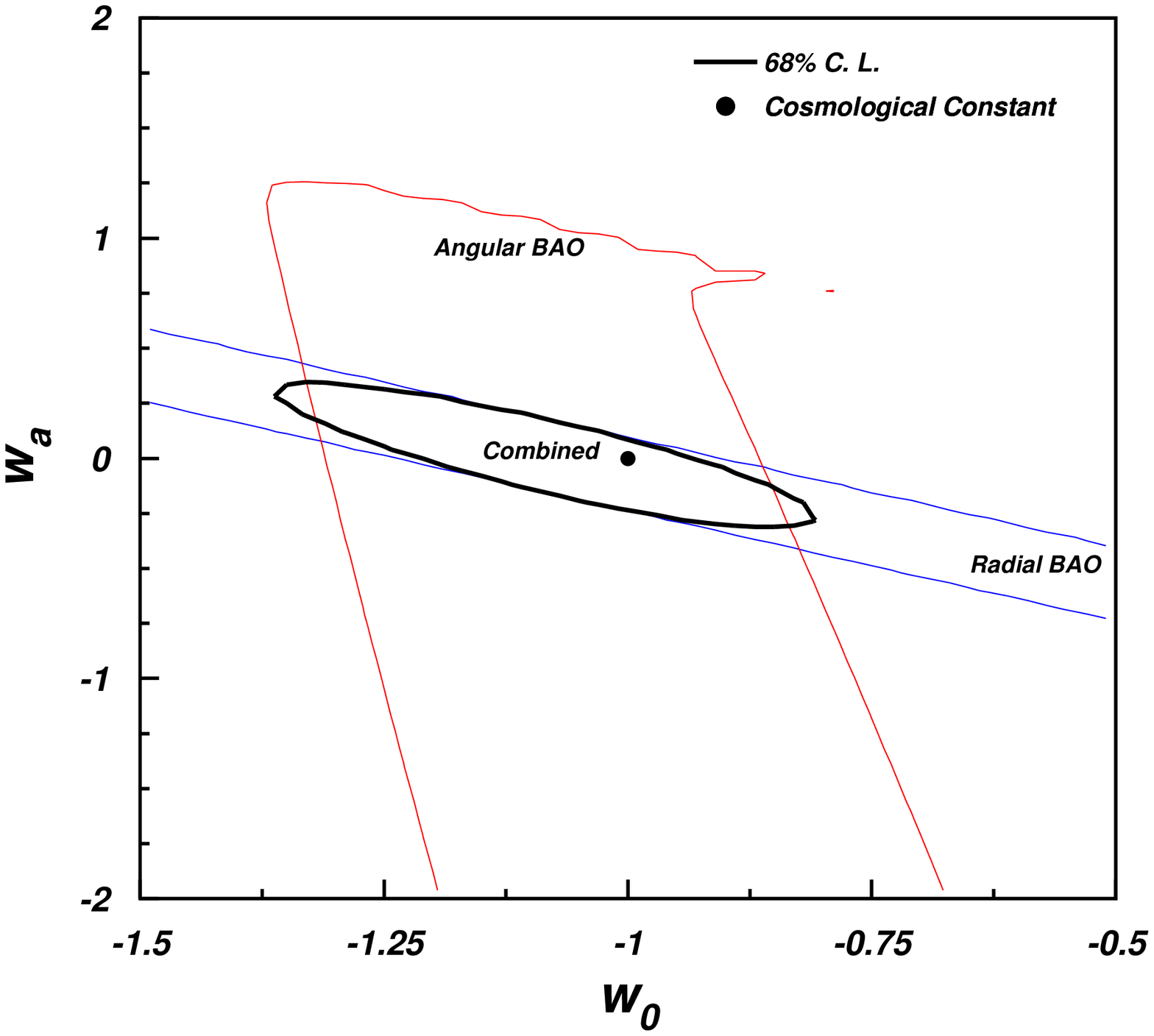}  \\
\end{tabular}
\vspace{-1.0truecm}
\caption{Left: The contribution of our determination $\theta_{BAO}(z=0.55)$, to the dark energy equation of state constraints. The bold contour is the full combination, while the light contour is obtained when excluding our measurement at $z=0.55$. The contribution 
         of this new measurement is not negligible, improving by 5\% the Figure of Merit. Right: Contribution 
         to the cosmological constraints from the angular (light vertical contour) and radial (light 
         diagonal contour) BAO scale measurements. Both determinations are 
         complementary, and the combination (bold contour) breaks the degeneracy, making the 
         constraints more precise. \label{fig:comparisons}}
\end{figure*}

The influence of our measurement in the cosmological constraints is shown in Figure~\ref{fig:comparisons} (Left), where we compare the constraints at 68\% C. L. for all the data (bold) with the constraints obtained when the point of this study ($z=0.55$) is excluded (light). The contribution of the new point is 
not negligible, since its inclusion makes the constraints more precise. In particular, the
figure of merit, defined as the inverse of the area within contours improves by 5\%. Moreover, we have studied how the angular and the radial BAO scale measurements contribute to the cosmological constraints. This is depicted in 
Figure~\ref{fig:comparisons} (Right), where the light vertical contour corresponds to the
angular BAO scales and the light diagonal contour corresponds to
the radial BAO scales. The bold contour are the full combination at 68\% C. L., also
presented in Figure~\ref{fig:w0wa} (Right). Both determinations are complementary and
the combination clearly breaks degeneracies between cosmological parameters, enhancing
the sensitivity.

Some comments about the detection of the radial BAO are in order, since
there are some claims saying that there is no convincing evidence of
the detection  \citep{2010ApJ...719.1032K}. For them, detection means a preference for a model with BAO as compare to a model without BAO. Nevertheless, we also consider the
appreciation given in \cite{2011MNRAS.tmpL.211C},
who argue that it is not justified to use any of the current BAO measurements (radial or not) to do a clear 
model selection. Instead, \cite{2011MNRAS.tmpL.211C} show that BAO measurements  can  be used for parameter fitting within ωCDM models, which is what we present here.

\section{Conclusions}
\label{sec:conclu}

We have measured the BAO signal in the distribution of galaxies at the redshift interval $0.5 < z < 0.6$, using a LRG sample selected from the SDSS DR7 photometric
catalog. The BAO signal is detected with a statistical significance of 2.3 standard
deviations, as expected from the characteristics of the 
sample \citep{rsd_paper}. The result is in agreement with the WMAP7 
cosmology, we find $\theta_{BAO} (z=0.55) = (3.90 \pm 0.38)^{\circ}$ for the angular BAO scale, including systematic errors. This is the first direct measurement of the BAO signal in angular space, and complements previous three-dimensional averaged and radial BAO measurements, making a significant contribution to break degeneracies between cosmological parameters. Combining with previous BAO measurements, this translates into
$\omega = -1.03 \pm 0.16$ for the  equation of state parameter of the dark energy, and
$\Omega_M = 0.26 \pm 0.04$ for the matter density.

Our relative error determination in the angular BAO scale is 
$9 \%$. This is larger than the $6.5\%$ uncertainty found by \cite{2007MNRAS.378..852P}, the only previous photometric determination of the BAO scale. This difference is not totally surprising, since our method to recover the BAO position only 
uses the information around the BAO peak, while \cite{2007MNRAS.378..852P} includes all 
scales with $k>0.3$. The impact and analysis of systematic effects is also quite 
different. For example, the above error in Padmanabhan does not include the photometric
redshift uncertainties, while this effect dominates our systematic error budget 
(see Table\ref{tab:sys}). 

We have also studied the sensitivity of all BAO measurements obtanied up to 
date, to a time varying dark energy equation of state, finding $\omega_a = 0.06 \pm 0.22$ when we fix all the other parameters to the WMAP7 cosmolgy. Therefore, the data is well described by a dark energy that behaves as a cosmological constant.

This study is a confirmation of the feasibility to obtain precise cosmological
constrains using photometric redshift surveys. This allows wider and deeper
galaxy samples, with the caveat of a less precise redshift. However, the smaller
precision in the redshift determination is compensated with the higher statistics, allowing a precise measurements of cosmological parameters. 

\section*{Acknowledgements}
\label{sec:acknowledgements}

Funding for the SDSS and SDSS-II has been provided by the 
Alfred P. Sloan Foundation, the Participating Institutions, the 
National Science Foundation, the U.S. Department of Energy, the 
National Aeronautics and Space Administration, the Japanese 
Monbukagakusho, the Max Planck Society, and the Higher Education 
Funding Council for England. The SDSS Web Site is 
http://www.sdss.org/.

    The SDSS is managed by the Astrophysical Research 
Consortium for the Participating Institutions. The Participating 
Institutions are the American Museum of Natural History, Astrophysical 
Institute Potsdam, University of Basel, University of Cambridge, Case 
Western Reserve University, University of Chicago, Drexel 
University, Fermilab, the Institute for Advanced Study, the Japan 
Participation Group, Johns Hopkins University, the Joint Institute 
for Nuclear Astrophysics, the Kavli Institute for Particle 
Astrophysics and Cosmology, the Korean Scientist Group, the Chinese 
Academy of Sciences (LAMOST), Los Alamos National Laboratory, the 
Max-Planck-Institute for Astronomy (MPIA), the Max-Planck-Institute 
for Astrophysics (MPA), New Mexico State University, Ohio State 
University, University of Pittsburgh, University of 
Portsmouth, Princeton University, the United States Naval 
Observatory, and the University of Washington.
 
We thank the Spanish Ministry of Science and Innovation (MICINN) for
funding support through grants AYA2009-13936-C06-01, 
AYA2009-13936-C06-03, AYA2009-13936-C06-04 and through the Consolider 
Ingenio-2010 program, under project CSD2007-00060.

We thank Carlos Cunha for his valuable comments about the photometric
redshift.


\begin{thebibliography}{}

\bibitem[\protect\citeauthoryear{{Abazajian}, {Adelman-McCarthy},
  {Ag{\"u}eros}, {Allam}, {Allende Prieto}, {An}, {Anderson}, {Anderson},
  {Annis}, {Bahcall} \& et al.}{{Abazajian} et~al.}{2009}]{2009ApJS..182..543A}
{Abazajian} K.~N.,  {Adelman-McCarthy} J.~K.,  {Ag{\"u}eros} M.~A.,  {Allam}
  S.~S.,  {Allende Prieto} C.,  {An} D.,  {Anderson} K.~S.~J.,  {Anderson}
  S.~F.,  {Annis} J.,  {Bahcall} N.~A.,    et al. 2009, ApJS, 182, 543

\bibitem[\protect\citeauthoryear{{Blake}, {Collister}, {Bridle} \&
  {Lahav}}{{Blake} et~al.}{2007}]{2007MNRAS.374.1527B}
{Blake} C.,  {Collister} A.,  {Bridle} S.,    {Lahav} O.,  2007, MNRAS, 374,
  1527

\bibitem[\protect\citeauthoryear{{Cabr{\'e}} \& {Gazta{\~n}aga}}{{Cabr{\'e}} \&
  {Gazta{\~n}aga}}{2011}]{2011MNRAS.tmpL.211C}
{Cabr{\'e}} A.,  {Gazta{\~n}aga} E.,  2011, MNRAS, 412, 98

\bibitem[\protect\citeauthoryear{{Cabr{\'e}}, {Gazta{\~n}aga}, {Manera},
  {Fosalba} \& {Castander}}{{Cabr{\'e}} et~al.}{2006}]{2006MNRAS.372L..23C}
{Cabr{\'e}} A.,  {Gazta{\~n}aga} E.,  {Manera} M.,  {Fosalba} P.,
  {Castander} F.,  2006, MNRAS, 372, 23

\bibitem[\protect\citeauthoryear{{Crocce}, {Cabre} \& {Gazta{\~n}aga}}{{Crocce}
  et~al.}{2010}]{2010arXiv1004.4640C}
{Crocce} M.,  {Cabre} A.,    {Gazta{\~n}aga} E.,  2010, ArXiv e-prints,
  1004.4640 [astro-ph]

\bibitem[\protect\citeauthoryear{{Crocce} et~al.,}{{Crocce}
  et~al.}{2011}]{rsd_paper}
{Crocce} M.,  et~al., 2011, ArXiv e-prints, 1104.5236 [astro-ph]

\bibitem[\protect\citeauthoryear{{Cunha}, {Lima}, {Oyaizu}, {Frieman} \&
  {Lin}}{{Cunha} et~al.}{2009}]{2009MNRAS.396.2379C}
{Cunha} C.~E.,  {Lima} M.,  {Oyaizu} H.,  {Frieman} J.,    {Lin} H.,  2009,
  MNRAS, 396, 2379

\bibitem[\protect\citeauthoryear{{Eisenstein} et~al.,}{{Eisenstein}
  et~al.}{2001}]{2001AJ....122.2267E}
{Eisenstein} D.~J.,  et~al., 2001, AJ, 122, 2267

\bibitem[\protect\citeauthoryear{{Eisenstein} et~al.,}{{Eisenstein}
  et~al.}{2005}]{2005ApJ...633..560E}
{Eisenstein} D.~J.,  et~al., 2005, ApJ, 633, 560

\bibitem[\protect\citeauthoryear{{Gazta{\~n}aga}, A. \& {Hui}}{{Gazta{\~n}aga}
  et~al.}{2009}]{2009MNRAS.399.1663G}
{Gazta{\~n}aga} E.,  A. C.,    {Hui} L.,  2009, MNRAS, 399, 1663

\bibitem[\protect\citeauthoryear{{Kaiser}, {Tonry} \& {Luppino}}{{Kaiser}
  et~al.}{2000}]{2000PASP..112..768K}
{Kaiser} N.,  {Tonry} J.~L.,    {Luppino} G.~A.,  2000, PASP, 112, 768

\bibitem[\protect\citeauthoryear{{Kazin}, {Blanton}, {Scoccimarro}, {McBride}
  \& {Berlind}}{{Kazin} et~al.}{2010}]{2010ApJ...719.1032K}
{Kazin} E.~A.,  {Blanton} M.~R.,  {Scoccimarro} R.,  {McBride} C.~K.,
  {Berlind} A.~A.,  2010, ApJ, 719, 1032

\bibitem[\protect\citeauthoryear{{Kazin} et~al.,}{{Kazin}
  et~al.}{2010}]{2010ApJ...710.1444K}
{Kazin} E.~A.,  et~al., 2010, ApJ, 710, 1444

\bibitem[\protect\citeauthoryear{{Komatsu} et~al.,}{{Komatsu}
  et~al.}{2011}]{2011ApJS..192...18K}
{Komatsu} E.,  et~al., 2011, ApJS, 192, 18

\bibitem[\protect\citeauthoryear{{Landy} \& {Szalay}}{{Landy} \&
  {Szalay}}{1993}]{1993ApJ...412...64L}
{Landy} S.~D.,  {Szalay} A.~S.,  1993, ApJ, 412, 64

\bibitem[\protect\citeauthoryear{{Lima}, {Cunha}, {Oyaizu}, {Frieman}, {Lin} \&
  {Sheldon}}{{Lima} et~al.}{2008}]{2008MNRAS.390..118L}
{Lima} M.,  {Cunha} C.~E.,  {Oyaizu} H.,  {Frieman} J.,  {Lin} H.,    {Sheldon}
  E.~S.,  2008, MNRAS, 390, 118

\bibitem[\protect\citeauthoryear{{Oyaizu}, {Lima}, {Cunha}, {Lin}, {Frieman} \&
  {Sheldon}}{{Oyaizu} et~al.}{2008}]{2008ApJ...674..768O}
{Oyaizu} H.,  {Lima} M.,  {Cunha} C.~E.,  {Lin} H.,  {Frieman} J.,    {Sheldon}
  E.~S.,  2008, ApJ, 674, 768

\bibitem[\protect\citeauthoryear{Padmanabhan et~al.,}{Padmanabhan
  et~al.}{2007}]{2007MNRAS.378..852P}
Padmanabhan N.,  et~al., 2007, MNRAS, 378, 852

\bibitem[\protect\citeauthoryear{{Percival}, {Cole}, {Eisenstein}, {Nichol},
  {Peacock}, {Pope} \& {Szalay}}{{Percival} et~al.}{2007}]{2007MNRAS.381.1053P}
{Percival} W.~J.,  {Cole} S.,  {Eisenstein} D.~J.,  {Nichol} R.~C.,  {Peacock}
  J.~A.,  {Pope} A.~C.,    {Szalay} A.~S.,  2007, MNRAS, 381, 1053

\bibitem[\protect\citeauthoryear{{Percival} et~al.,}{{Percival}
  et~al.}{2010}]{2010MNRAS.401.2148P}
{Percival} W.~J.,  et~al., 2010, MNRAS, 401, 2148

\bibitem[\protect\citeauthoryear{{Reid} et~al.,}{{Reid}
  et~al.}{2010}]{2010MNRAS.404...60R}
{Reid} B.~A.,  et~al., 2010, MNRAS, 404, 60

\bibitem[\protect\citeauthoryear{{Riess} et~al.,}{{Riess}
  et~al.}{2011}]{2011ApJ...730..119R}
{Riess} A.~G.,  et~al., 2011, ApJ, 730, 119

\bibitem[\protect\citeauthoryear{{S{\'a}nchez}, {Crocce}, {Cabr{\'e}}, {Baugh}
  \& {Gazta{\~n}aga}}{{S{\'a}nchez} et~al.}{2009}]{2009MNRAS.400.1643S}
{S{\'a}nchez} A.~G.,  {Crocce} M.,  {Cabr{\'e}} A.,  {Baugh} C.~M.,
  {Gazta{\~n}aga} E.,  2009, MNRAS, 400, 1643

\bibitem[\protect\citeauthoryear{{S{\'a}nchez} et~al.,}{{S{\'a}nchez}
  et~al.}{2011}]{2011MNRAS.411..277S}
{S{\'a}nchez} E.,  et~al., 2011, MNRAS, 411, 277

\bibitem[\protect\citeauthoryear{{Schlegel} et~al.,}{{Schlegel}
  et~al.}{2007}]{2007AAS...21113229S}
{Schlegel} D.~J.,  et~al., 2007, in American Astronomical Society Meeting
  Abstracts Vol.~38 of Bulletin of the American Astronomical Society,
  {SDSS-III: The Baryon Oscillation Spectroscopic Survey (BOSS)}.
p. 132.29

\bibitem[\protect\citeauthoryear{{Schlegel} et~al.,}{{Schlegel}
  et~al.}{2009}]{2009arXiv0904.0468S}
{Schlegel} D.~J.,  et~al., 2009, ArXiv e-prints, 0904.0468 [astro-ph]

\bibitem[\protect\citeauthoryear{{The Dark Energy Survey Collaboration}}{{The
  Dark Energy Survey Collaboration}}{2005}]{2005astro.ph.10346T}
{The Dark Energy Survey Collaboration} 2005, ArXiv e-prints, astro-ph/0510346

\bibitem[\protect\citeauthoryear{{Thomas}, {Abdalla} \& {Lahav}}{{Thomas}
  et~al.}{2011}]{2010arXiv1011.2448T}
{Thomas} S.~A.,  {Abdalla} F.~B.,    {Lahav} O.,  2011, MNRAS, 412, 1669

\bibitem[\protect\citeauthoryear{{Tyson}, {Wittman}, {Hennawi} \&
  {Spergel}}{{Tyson} et~al.}{2003}]{2003NuPhS.124...21T}
{Tyson} J.~A.,  {Wittman} D.~M.,  {Hennawi} J.~F.,    {Spergel} D.~N.,  2003,
  Nuclear Physics B Proceedings Supplements, 124, 21

\bibitem[\protect\citeauthoryear{{York} et~al.,}{{York}
  et~al.}{2000}]{2000AJ....120.1579Y}
{York} D.~G.,  et~al., 2000, AJ, 120, 1579

\end{thebibliography}

\end{document}